\documentclass[proceedings, preprint]{rmaa}

\usepackage{wasysym}

\usepackage{paralist}

\usepackage{psfrag,color}

\newcommand{\Teff}{$T_{\rm eff}$}
\newcommand{\logg}{$\log g$}

\SetYear{2007}
\SetConfTitle{XII Reunion Latinoamericana}

\title{Model stars for the modelling of galaxies: $\alpha$-enhancement in
stellar populations models} 

\author{P. Coelho\altaffilmark{1}}

\altaffiltext{1}{Marie Curie Felllow, Institut d'Astrophysique, CNRS, 
Universit\'e Pierre et Marie Curie, 98 bis Bd Arago, 75014 Paris, 
France (pcoelho@iap.fr).}

\shortauthor{Coelho}
\shorttitle{Model Stars \& Model Galaxies}

\listofauthors{P. R. T. Coelho}
\indexauthor{Coelho, P. R. T.}

\abstract{
Stellar population (SP) models are an essential tool to understand the
observations of galaxies and clusters. One of the main ingredients of
a SP model is a library of stellar spectra, and both empirical and theoretical
libraries can been used for this purpose. Here I will start by giving a short
overview of the pros and cons of using theoretical libraries, i.e.
model stars, to produce our galaxy models. Then I will 
address the question on how theoretical libraries can be used to model stellar
populations, in particular to explore the effect of $\alpha$-enhancement
on spectral observables.
}

\resumen{}

\addkeyword{Galaxies: Abundances}
\addkeyword{Galaxies: Stellar content}
\addkeyword{Stars: Abundances}
\addkeyword{Stars: Atmospheres}

\begin{document}
\maketitle

\section{Introduction}
\label{s:intro}
To understand how galaxies form and evolve is one of the long-standing questions
in astronomy. One of the ways to address that question  
is to study the stellar content of the galaxies through the use of stellar population (SP) synthesis techniques, i.e., the modelling of the spectral energy distribution (SED) emitted by evolving stellar populations \citep[see][]{tinsley80}. An evolutionary SP model has 
two main ingredients: a set of stellar evolutionary models 
(tracks or isochrones) that will predict how the
stars are 
distributed in the HR diagram, and a library of stellar
observables (e.g. colours, spectra, spectral indices, etc.) that coupled
to the evolutionary models, will be used to predict
the colours or spectra of a stellar population. 
By computing models for several choices of SP parameters 
(age {\it t}, metallicity {\it Z}, initial mass function {\it IMF}, etc.), 
it is possible to produce a grid of SP models that can be used to study a
variety of systems, from early-type galaxies and spiral bulges to star 
forming galaxies at different redshifts. 

A keystone to extract the star formation history {\it SFH} of unresolved stellar 
systems is through the analysis of their chemical abundance pattern. 
The chemical pattern is a tracer of the history of star formation  
because different elements are produced at different timescales during the evolution 
of a galaxy (e.g. Greggio \& Renzini 1983, Matteucci \& Greggio 1986).  
In fact, the $\alpha$-elements 
over Fe abundances ([$\alpha$/Fe]) is largely used to constraint 
the formation timescale of a galaxy,
since the $\alpha$-elements (O, Ne, Mg, Si, S, Ca, and Ti) 
are released early in the evolution 
of the galaxy by Type II supernovae (SNe II), while Fe is mainly produced by 
Type Ia supernovae (SNe Ia) on longer time-scales.
Galaxies that had undergone rapid star formation present [$\alpha$/Fe] values
larger than the ratios found in the Milky Way disc, reason why SP
models built with stars of the solar neighbourhood cannot reproduce
the locus of more massive galaxies in Mg versus Fe indices plots
\citep[e.g.][]{worthey+92}.
Therefore, large efforts have being made by different groups in 
computing SP models with different [$\alpha$/Fe]. 
I will focus on the role that the theoretical stellar 
spectra computations have 
in modelling populations with different [$\alpha$/Fe].

\section{Theoretical stellar libraries}
\label{s:stellib}
The library of stellar spectra used in the SP models can be either
empirical or theoretical, a choice which is subject to debate. There 
are plenty of models in 
literature using both empirical (e.g. 
\citealt{vazdekis99}, the visible range of \citealt{bc03} and
\citealt{PEGASE-HR}) or theoretical libraries (e.g. UV and IR range of \citealt{bc03}, 
\citealt{delgado+05}, \citealt{maraston05}, \citealt{zhang+05}). 
An empirical library is based on observed stellar spectra covering as much as
possible the 
parameters \Teff, \logg, Z. 
It is not a simple task to assemble a library 
that simultaneously features high S/N, good flux calibration, large wavelength coverage, 
high-resolution and accurately derived stellar 
parameters, but great improvements have been made 
\citep[e.g.][]{ELODIE,INDOUS,MILES}. The major drawback of an empirical
library is that such high quality observations are limited to the closest stars, and
therefore the coverage of the HR diagram and the stellar abundances 
are biased towards the solar neighbourhood.

A theoretical stellar library (also called synthetic stellar library)
is based on model atmospheres and 
atomic and molecular line list. A {\it model atmosphere} is the run
of temperature, gas, electron and radiation pressure, 
convective velocity and flux, and more generally, of all relevant quantities
as a function of some depth variable (geometrical, or optical depth
at some special frequency, or column mass).
The {\it synthetic spectrum}, or {\it flux distribution} is the 
emergent flux computed based on a model atmosphere, and is required for comparison
with observations.
Theoretical libraries have the advantadge of covering the parameter space in 
\Teff, \logg, and abundances at will. Moreover, a synthetic
star has very-well defined atmospheric parameters and infinite S/N, and covers a larger wavelength coverage
with (possibly) higher resolution than a single observed spectrum. 
To compute a large synthetic library can be demanding 
in terms of computational time, but it is usually feasible. There is a caveat though: being based on our knowledge
on stellar atmospheres and databases of atomic and molecular transitions, 
those libraries are
limited by the approximations and (in)accuracies of their underlying models, and in fact
we are not able
to reproduce accurately all spectral types
\citep[see e.g.][]{gustafsson+07proc}. There is a lot
to be done yet in terms of including in a
realistic way effects of 3D hydrodynamics, asphericity, N-LTE, winds, non-radiative heating, chromospheric contribution, etc. Moreover,
the databases of atomic and molecular transitions provide relatively few lines with highly accurate oscillator strengths and broadening parameters,
besides being often incomplete 
\citep[see the progress report by][]{kurucz06}.
An additional point of confusion for the SP modeller (and user) is that theoretical
libraries can be categorised into two groups: 
\begin{itemize}
\item libraries computed with a good treatment of the line-blanketing, therefore being
good for spectrophotometric
predictions and low-resolution studies \citep[e.g.][]{ATLASODFNEW,MARCS03,PHOENIX05}; 
\item libraries which are computed with shorter, fine-tuned, empirically calibrated
atomic and molecular line lists, being the ones more appropriate 
for high-resolution studies 
\citep[e.g.][]{peterson+01,uvblue05,coelho+05}.   
But in general they are not suitable to predict colours due to missing 
line-blanketing. 
\end{itemize}

The ability of some of the recent theoretical libraries in reproducing 
observations was recently assessed in \citet{MC07}.
Concerning the libraries
for low-resolution studies, broad-band colours predictions 
by three libraries were compared to an empirical
UBVRIJHK calibration.
Models can reproduce with reasonable accuracy the stellar colours for a fair interval
in effective temperatures and gravities, but there are some problems with
U-B and B-V colours
and very cool stars in general (V-K $\apprge$ 3). The results for the B-V and V-I colours
are illustrated in Fig. 1.
As for the libraries aimed at high-resolution studies, 
\citet{MC07} analysed the performance
of three libraries by comparing their
predictions for spectral indices to measurements given by empirical libraries.
Fig. 2 shows the results for three indices against the empirical library 
MILES \citep{MILES}. 
In general the libraries present
similar behaviours and systematic deviations. 
In particular, the lists of atomic and molecular
lines need further improvement, specially in the blue region of the
spectrum and 
for the stars with {\Teff} $\apprle$ 4500K.

\begin{figure}[!t]
\center
\includegraphics[width=5cm]{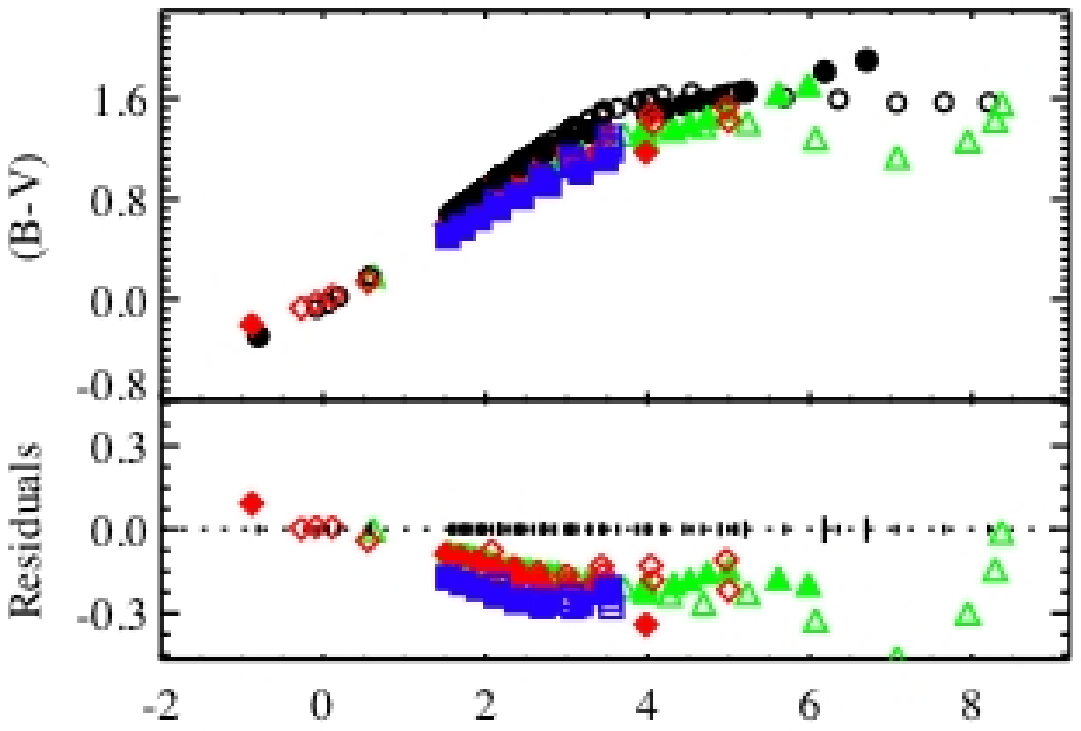}
\includegraphics[width=5cm]{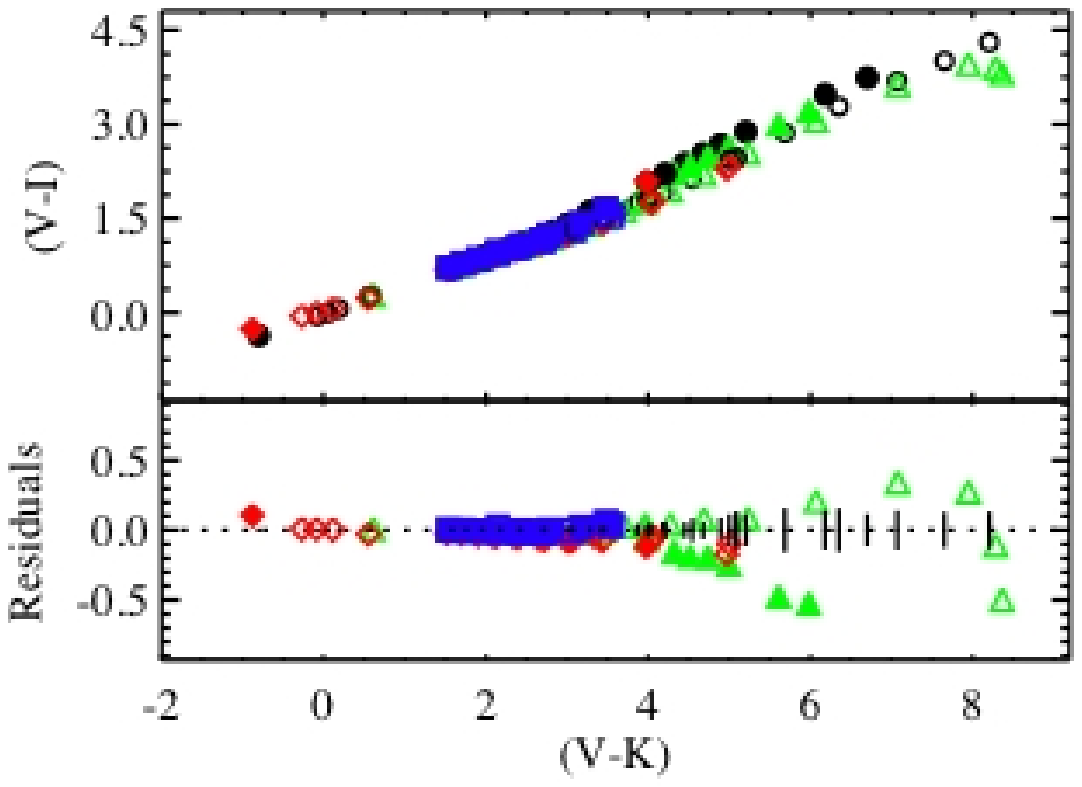}
\caption{
Comparison between the colours predicted by synthetic flux distributions 
and the colour-temperature relation by \citet{worthey_lee06astroph}, for stars
representative of a 10 Gyr population. 
Red-diamonds correspond to ATLAS9 \citep{ATLASODFNEW}, 
green-triangles to PHOENIX
\citep{PHOENIX05} and the blue squares to MARCS models \citep{MARCS03}.
Filled and open symbols represent dwarf and giant stars respectively. 
Filled black-circles
are the values expected from the empirical relation.
The bottom of each colour plot shows the residuals (difference between models 
predictions and empirical calibration; the thin black vertical lines
are the error bars of the empirical calibration).
Adapted from \citet{MC07}.}
\label{fig:simple}
\end{figure}

\begin{figure}[!t]
\includegraphics[width=\columnwidth]{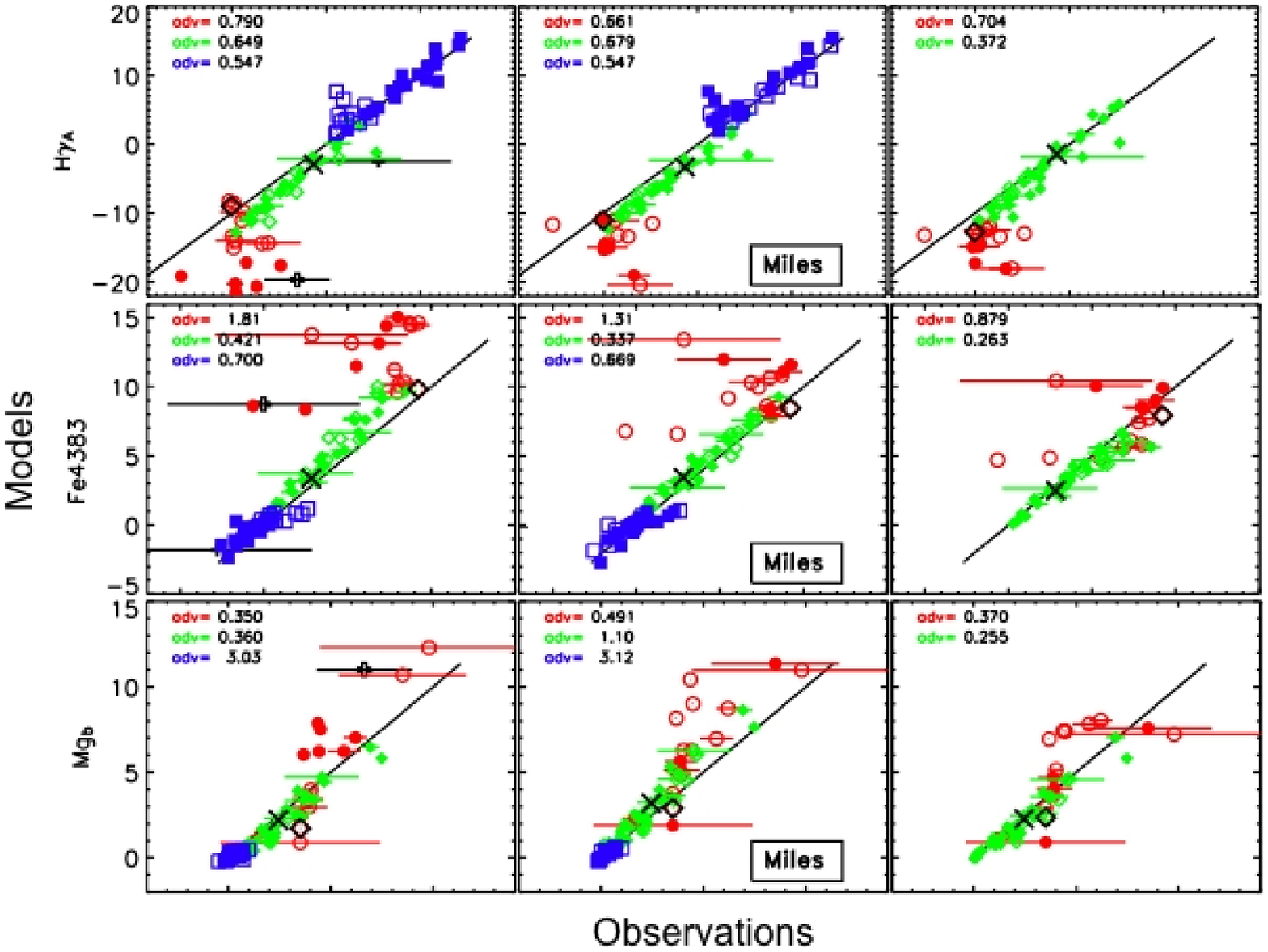}
\caption{Comparison between the predictions of three synthetic libraries
in the $y$-axis (\citealt{martins+05} in the left-hand column; \citealt{munari+05} in the
middle colum; \citealt{coelho+05} in the right-hand column)
and indices measured in the MILES empirical library in the $x$-axis, 
for three spectral indices indicated in the figure.
Blue squares are stars with \Teff $<$ 7000K, green diamonds for
4500K $<$ \Teff $\leq$ 7000K and red circles for 
\Teff $\leq$ 4500K (filled and open symbols are dwarf and giant stars respectively). 
The solid line is the one-to-one relation. The thick black symbols 
represent a Sun-like dwarf (diamond) and an Arcturus-like giant (cross).
Adapted from \citet{MC07}.}
\label{fig:simple}
\end{figure}

\section{Including $\alpha$-enhancement effects in stellar population models}
\label{s:stelpop}

As mentioned in the Introduction, the spectra of the stars carry the chemical signatures that are imprints of the SFH of the galaxy they belong to.
Hence, and despite the current limitations of the theoretical libraries as 
mentioned in \S~2,
we cannot rely solely on the empirical libraries if we intend to reproduce 
the spectra of galaxies which have undergone a star formation different from 
that of the solar neighbourhood. In fact, there is more than one way
of computing SP models with variable [$\alpha$/Fe] values.

\subsection{The first SP models with non-solar patterns} 

The first attempts to model SPs beyond the solar-scaled
pattern date back to mid-90s.
\citet{barbuy94} used 
synthetic stellar spectra to 
predict the Mg$_2$ index for ages and abundances typical 
of those of bulge globular clusters. 
\citet{weiss+95}, 
using observed spectra of bulge stars
and for the first time including $\alpha$-enhanced evolutionary tracks for a high 
metallicity population, computed 
models for Mg$_2$ and $<$Fe$>$ indices. 
Alternatively to the use of spectra, \citet{borges+95}
adopted empirical fitting functions\footnote{
The {\it fitting functions} describe how
stellar spectral indices vary as a function of the stellar parameters 
\Teff, \logg, [Fe/H] or [Z/H] and sometimes, abundance [X/Fe] of 
another element relative to iron.}
with explicit dependence on [Mg/Fe] or [Na/Fe] to model SP indices with variable
Fe, Mg and Na abundances.
Other authors further explored the conclusions and methods of the
pioneer efforts, e.g., \citet{idiart+96, vazdekis+97, tantalo+98}. Those first
models were limited to a few indices or to a restricted range of SP parameters.

\subsection{Fitting functions \& Sensitivity tables}
The study by  
\citet{TB95} opened a way to produce non-solar-scaled models 
for {\it all} the Lick/IDS indices. Based on synthetic spectra calculations, they
quantified how each of the Lick/IDS indices changes with variations of
individual chemical elements. These {\it sensitivity tables} (also called {\it response functions}),
were computed for three combinations of \Teff\ and \logg, which correspond to a main-sequence dwarf, a turn-off star and a red-giant star of a 5 Gyr-old stellar 
population. 

By combining empirical and theoretical information,
\citet{trager+00a} proposed a method to use the sensitivity tables by \citet{TB95} to
apply corrections to the empirical fitting functions by \citet{worthey+94}, in order to produce SP model indices with arbitrary compositions
\footnote{Alternatively, theoretical fitting 
functions with explicit dependence on $\alpha$/Fe \citep{barbuy+03} can be 
employed to compute $\alpha$-enhanced models \citep{oliveira+05}, but the 
theoretical fitting functions were limited to a few Lick/IDS indices.}. The method was  extended by \citet{TMB03} who published a large set of SP models for all Lick/IDS indices and variable values of [$\alpha$/Fe], [$\alpha$/Ca] and
[$\alpha$/N]. Until the present time, this technique is the most widely used to
model SPs indices with variable abundances patterns, and both model indices and
ingredients (sensitivity tables and fitting functions)
are constantly updated
\citep[e.g.][]{houdashelt+02,tantalo_chiosi04, korn+05,lee+worthey05,
annibali+07,martinhernandez+07,schiavon07}.

Important advancements in our knowledge of the chemical composition
of galaxies have been achieved through the use of those models 
\citep[e.g.][]{kuntschner+01,PS02,proctor+04,thomas+05,kelson+06,smith+06,delarosa+07}. 
Nevertheless this approach is subject to some limitations: 
\begin{itemize}
\item the responses of three synthetic stars are used to correct all stars in an SP model and the impact of this approximation in the accuracy of the model predictions is uncertain; 
\item only Lick/IDS indices can be modelled. Hence 
full-spectrum high-resolution models such as \citet{vazdekis99,bc03,PEGASE-HR} 
remain constrained to the solar neighbourhood chemical pattern, hampering
the use of full-spectrum fitting techniques, colours or non-Lick indices
in interpreting observations.
\item the effect of the abundance variations on the isochrones  
is often not included in a consistent way.
\end{itemize}

\subsection{Fully theoretical approach}

Recent progress in the modelling of high-resolution stellar spectra opened the
door to a new kind of models, in which the effects of abundance variations can be
studied at any wavelength. This is enabled by the publication of several 
libraries of theoretical, high-resolution stellar spectra for both solar-scaled
and $\alpha$-enhanced chemical mixtures \citep{PHOENIX05,coelho+05,munari+05}. 

The first fully theoretical high-resolution SP models for an 
$\alpha$-enhanced mixture were presented in \citet{coelho+07}, who
used an improved version of the stellar library in \citet{coelho+05}
to compute SP models for solar-scaled and $\alpha$-enhanced compositions
for three values of iron abundance [Fe/H] and ages from
3 to 14 Gyrs. These models employ
newly computed stellar tracks \citep{weiss+07proc} 
with the same abundances as the stellar library.
For the $\alpha$-enhanced mixture it is adopted a flat-enhancement of 0.4 dex for the
abundance ratios of all the classical $\alpha$ elements (O, Mg, Si, S, Ca and Ti).
The impact of the spectral and evolutionary
effects is illustrated on the Figs.~3 and 4 for broad-band colours and spectral indices, respectively. The modelling of colours requires that both effects are taken into account,
as they often have comparable effects. As for the spectral indices, the spectral
effect is overall the dominant one, but the evolutionary effect can be non-negligible (e.g. Balmer indices and near-IR indices).

Another theoretical study at high-resolution is being developed 
by \citet[][Lee et al. in prep.]{lee+07proc}, 
based on the stellar evolutionary tracks by \citet{dotter+07} and
with the flexibility to explore the effect of abundance variations 
in an element-by-element basis.

The fully-theoretical method has the advantage of providing 
a larger coverage in wavelength at higher
resolution than the methods in \S~3.2, and of providing more accurate predictions 
given that the effect of the $\alpha$-enhancement on the stellar evolutionary 
tracks and spectral library
is included in a consistent way. The caveat of this approach is that given the 
limitations of the synthetic libraries mentioned in \S~2, these models are more 
affected by zero-point problems than semi-empirical methods. 
Hence they are not a straightforward
replacement for the models based on empirical libraries, being 
more suitable to differential analysis 
\footnote{There are interesting work on deriving the stellar population parameters in a differential way, avoiding zero-point
problems that affect both models and data. See e.g. \citet[][]{nelan+05,kelson+06}.}.

\begin{figure}[!t]
\includegraphics[width=\columnwidth]{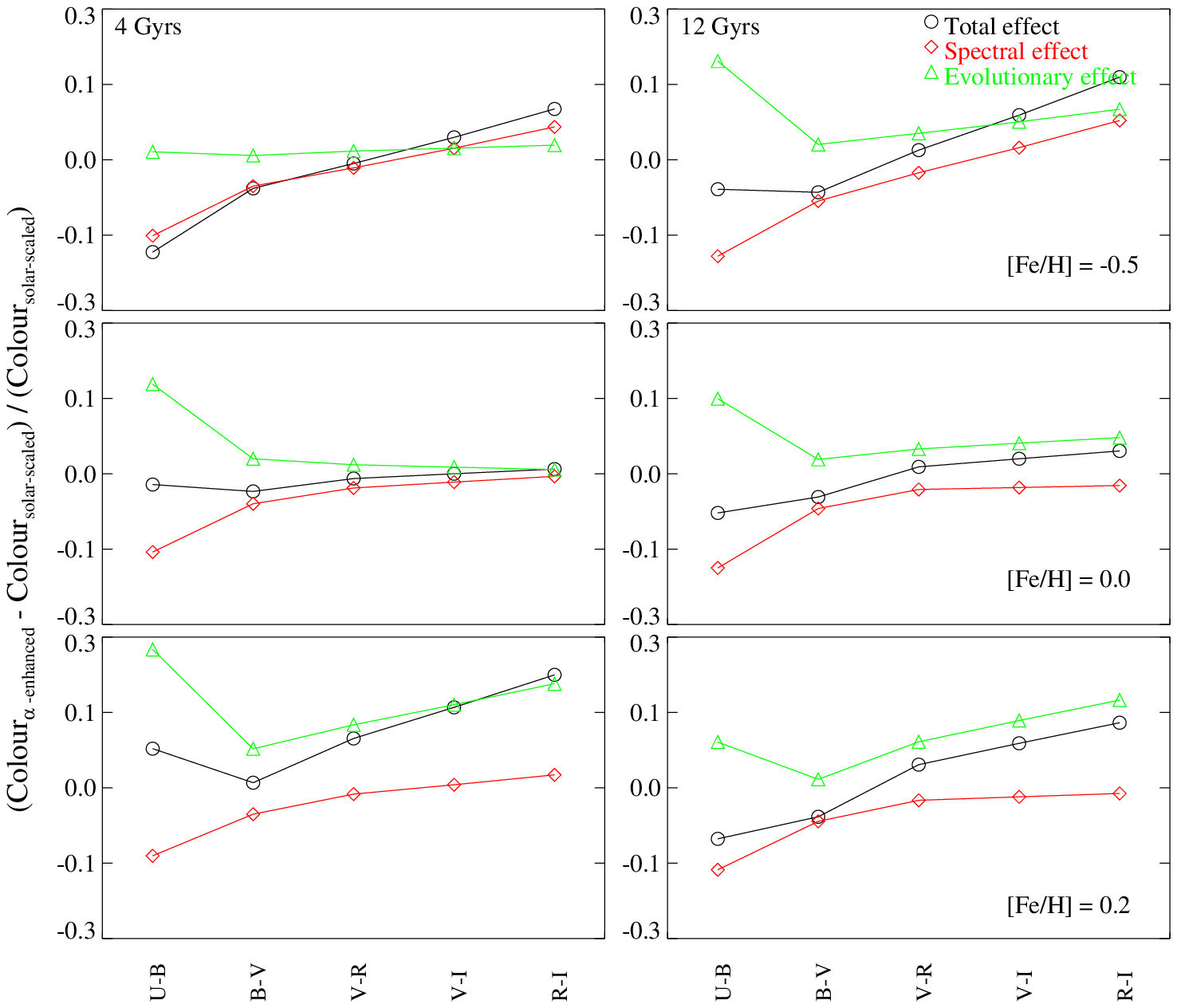}
\caption{Differences between the $\alpha$-enhanced predictions
and the solar-scaled ones, in units of the solar-scaled value, for broad-band colours 
shown on {\it x}-axis. 
Each row corresponds to a different [Fe/H] value, indicated in the figure. 
Left- and
righ-hand columns show the predictions for 4 Gyr and 12 Gyr SSPs respectively.
The green-triangles and red-diamonds lines correspond to the evolutionary effect and spectral effects respectively. 
The
black-circles line are the predictions when both effects are considered together. From \citet{coelho+07}.}
\label{fig:simple}
\end{figure}

\begin{figure}[!t]
\includegraphics[width=\columnwidth]{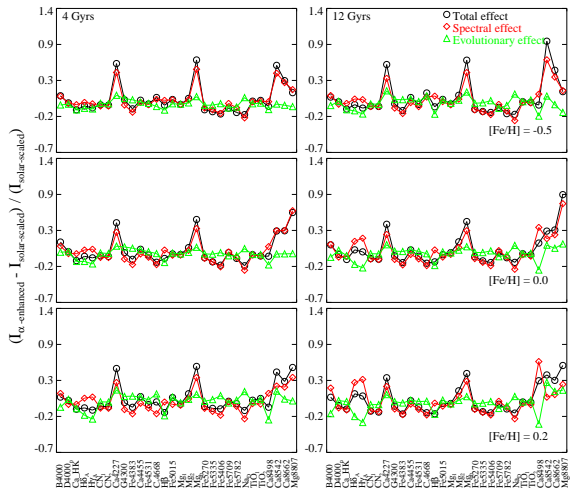}
\caption{Differences between the $\alpha$-enhanced predictions
and the solar-scaled ones, in units of the solar-scaled value, for several
spectral indices shown on {\it x}-axis.
Each row corresponds to a different [Fe/H] indicated in the figure. 
Left- and
right-hand columns show the predictions for 4 Gyr and 12 Gyr SSPs, respectively.
The green-triangles and red-diamonds lines correspond to the evolutionary effect and spectral effects 
respectively. The black-circles line are the predictions when both effects are considered 
together. From \citet{coelho+07}.}
\label{fig:simple}
\end{figure}

\subsection{Spectral corrections}

There is ongoing work that aim at combining the versatility of the method in
\S~3.3 in terms of exploring wider wavelength ranges at higher resolution,
and the accuracy of SP models 
based on empirical libraries.
Similarly to using response functions to correct indices given by fitting functions,
theoretical stellar libraries can be used to differentially correct SP models based
on empirical libraries. To my knowledge, there are two groups working on this approach.
\citet{prugniel+07proc} recently used the theoretical library by \citet{coelho+05}
to differentially correct the stars of the empirical library ELODIE \citep{ELODIE}. 
Two `semi-empirical' libraries were produced
with [Mg/Fe] = 0.0 and +0.4, and intermediate values were linearly interpolated as a
function of the mass ratio of Mg to Fe. This semi-empirical library can then 
be used to build SP models with variable [Mg/Fe].
Alternatively, the correction to the $\alpha$-enhanced pattern can be applied
{\it a posteriori}: SP models computed with theoretical stars 
are used to differentially modify SP models based on the
empirical library. An application of this method was presented in \citet{cervantes+07}, using newly computed theoretical stars to correct SP models based on the MILES
library (Vazdekis et al. in prep.).

\section{Concluding remarks}
\label{s:conclu}

Considerable efforts have being applied in the modelling of
SP models with variable chemical patterns. Following the pioneer efforts
mentioned in \S 3.1, the models that combined fitting functions and sensitivity tables
(\S 3.2) were a breakthrough, and they had an
important impact on our knowledge of the abundances in galaxies derived
by spectral indices.
Recently, fully-theoretical models (\S 3.3) started
to fill a gap which existed among the full spectral high-resolution SP models,
providing ways of modelling a much larger number of observables
and including spectral and evolutionary effects in a consistent way. 
And there are other methods being developed nowadays, like the empirical 
corrections mentioned in \S 3.4. 
Each of the methods has its strong and weak points
and we can expect future improvements in all of them. I think that there is not
a `best' approach. The user of stellar population models should choose
the model family that best suits his/her application, always 
keeping in mind the weakness and strengths of each approach. 
But we can be certain of one thing: one way or another,
we do need model stars in our galaxy models. 

{\it Acknowledgements}:
P. Coelho thanks the organisers for their kind hospitality and financial support.

\end{document}